\documentclass[useAMS,usenatbib]{mn2e}
\usepackage{psfig}

\title[SNII enrichment and the star cluster mass function]{SNII 
enrichment and the star cluster mass function}

\author[Simon\,P.\,Goodwin and
  B.\,E.\,J.\,Pagel]{Simon\,P.\,Goodwin$^1$\thanks{E-mail: 
Simon.Goodwin@astro.cf.ac.uk; bejp@sussex.ac.uk} and
  B.\,E.\,J.\,Pagel$^2$ \\
$^1$ Department of Physics \& Astronomy, Cardiff
University, 5 The Parade, Cardiff, CF24 3YB, UK \\
$^2$Astronomy Centre, University of Sussex, Falmer, Brighton, BN1 9QH,
UK}

\begin{document}

\date{}

\pagerange{\pageref{firstpage}--\pageref{lastpage}} \pubyear{2002}

\maketitle

\label{firstpage}

\begin{abstract}
Changing the form of the star cluster mass function (CMF) can effectively
change the upper end of the stellar initial mass function.  The yields
of from supernovae are very sensitive to the
mass of the progenitor star.  We show that by changing the parameters
of the CMF, it is possible to change the yields of oxygen and magnesium 
by a factor of $\sim 1.5$ and of metals in general by a factor of $\sim 1.8$.  

\end{abstract}


\section{Introduction}

Virtually all stars form in clusters (Clarke, Bonnell \& Hillenbrand 
2000; Lada \&  Lada
2003).  The type-II supernovae (SNII) of the most massive  stars in
these clusters are the main source of many heavy  elements, in
particular oxygen and $\alpha$-elements.  The yields of  heavy
elements from these SNII depend upon the mass of the  progenitor
(e.g. Tsujimoto et al. 1995; Woosley \& Weaver 1995;  Hoffman et
al. 1999; Limongi, Straniero \& Chieffi 2000; Rauscher et al. 2002; Heger  et
al. 2002).  Thus the yields of heavy elements will depend upon the
initial mass function (IMF) of the most massive stars in clusters.  Of
particular interest in the study of SNII products are oxygen and
magnesium that are both thought to be pure SNII products and which are
relatively easy to observe.

In large elliptical galaxies N/Fe, Na/Fe and Mg/Fe seem to be about  a
factor of 2 supersolar (Henry \& Worthey 1999), as do Mg/Fe and  O/Fe
in stars of the Galactic halo and Thick disk (Prochaska et al.
2000). In the Galactic bulge,  anomalies in ratios such as Mg/Ca
suggest changes in the initial mass  function (McWilliam \& Rich
2004), which could also play a role in  the $\alpha$/Fe ratios,
although their behaviour is  usually attributed to differences in
star formation history. In this investigation we are mainly concerned
with the overall oxygen and magnesium yields, which can be deduced
from the peak in the abundance distribution function of stars (Pagel
1997); this peak for Mg is evidently at a somewhat greater abundance
in the Galactic Bulge and the central regions of large elliptical
galaxies than the approximately solar value that it has in the solar
neighbourhood.

In a recent paper Kroupa \& Weidner (2003) showed that the (effective)
upper IMF depends upon the cluster mass function (CMF).  Low-mass
clusters are unlikely to contain very massive stars as they do not
have the gas reservoirs from which to form them.  A $10^2
M_\odot$ cluster cannot contain a star $> 100 M_\odot$ and is unlikely
to contain stars $>30 - 40 M_\odot$.  A $10^5 M_\odot$ cluster, on the
other hand, is very likely to contain some stars  $> 100 M_\odot$.
This idea is supported by evidence for a truncated IMF in small
clusters for stellar masses $> 10 M_\odot$ (Thilker et al. 2002).

In the Solar neighbourhood the CMF is a power-law with slope $\sim -2$
between $\sim 10^2$ and $10^4 M_\odot$ (Lada \& Lada 2003).  There is
evidence that the CMF varies with galaxy type (e.g. Kennicutt, Edgar
\& Hodge 1989; Thilker et al. 2002; Alonso-Herrero et al. 2002; Youngblood
\& Hunter 1999).

In this paper we discuss the effect on the yields of oxygen and
magnesium from SNII due to changing the CMF (and so effectively
changing the upper-end IMF).  In Section 2 we describe our method, in
Section 3 we present the results, and in Section 4 we discuss their
implications for heavy element yields in galaxies of different types
and masses.

\section{Method}

We select clusters from a cluster mass function (CMF) and then
populate these clusters with stars from a standard IMF.  Following 
Kroupa \& Weidner (2003) we select star cluster masses at random 
from a power-law cluster mass function of the form $N(M) \propto 
M^{-\beta}$ between lower and upper mass limits $M_{\rm low}$ and 
$M_{\rm up}$.

Each cluster is then populated with stars drawn at random from a
two-part Kroupa (2002) IMF of the form

\begin{displaymath}
N(M) \propto \left\{
\begin{array}{ll}
 M^{-1.3} & \,\,\,\, 0.08 < M/M_\odot < 0.5 \\ 
M^{-2.3} & \,\,\,\, 0.5 < M/M_\odot < 150 \\
\end{array} \right.
\end{displaymath}

\noindent such that the slope is Salpeter down to $0.5 M_\odot$ and
then flattens.  We ignore the brown dwarf regime as brown dwarfs,
although numerous, contribute very little mass to the cluster.

Stars are added to the cluster until the total mass of stars exceeds
the cluster mass.  If the sampling produces a total cluster mass more
than 2\% greater than the desired cluster mass, then the cluster is
completely repopulated.  In each Monte Carlo run, the random sampling
of clusters is continued until a total of  $10^9 M_\odot$ of stars
has been formed.

We record the numbers of stars greater than $8 M_\odot$ and the
enrichment from these stars.  Initially very massive stars $>
25 M_\odot$ are expected to lose significant amounts of material
through wind-driven mass-loss (e.g. Maeder 1992; Langer  \& Henkel
1995).  For example, the yields quoted by Maeder (1992; table 6) for
stars up to $\sim 25 M_\odot$ are very similar to those from Tsujimoto
et al. (1995) and Thielemann et al. (1996).  Above this mass, the
yields drop sharply due to the lower mass of the  eventual SNII.

\begin{figure*}
\centerline{\psfig{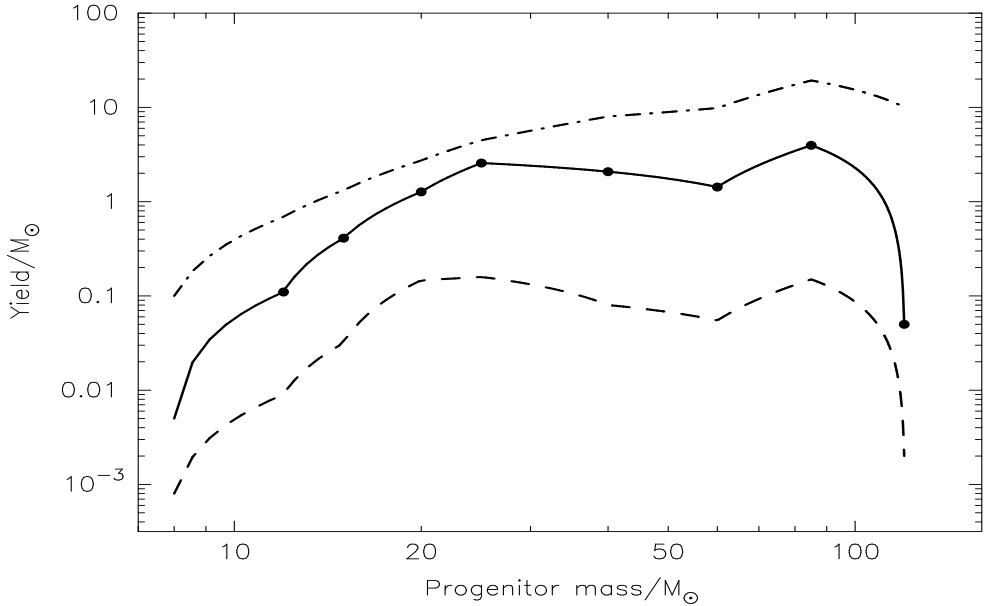}}
\caption{The yields of metals (dash-dot line), oxygen (full line) and 
magnesium (dashed line) against the initial stellar mass. Metal and
oxygen yields are taken from Maeder (1992) solar metallicity high
mass-loss models, and the tabulated masses from Maeder are marked on
the oxygen line.  Magnesium yields are taken from Tsujimoto et
al. (1995) and Thielemann et al. (1996) up to $25 M_\odot$, and are
then taken to be $10\%$ of the explosive yields of stars from Maeder
(1992).}
\label{fig:yields}
\end{figure*}

We take yields for oxygen (O) and metals (Z) from the combination of 
contents of
stellar winds and final ejecta given by Maeder (1992) for
solar-metallicity stars with strong mass loss (his Table 6).
We assume that the yield of magnesium follows Tsujimoto et al. (1995) 
up to $25M_\odot$, and that the Mg/O ratio does so above this. 
Between the tabulated  points
intermediate values are calculated by a linear  interpolation.  In
Fig.~\ref{fig:yields} we plot the O, Z and Mg yields against the initial
stellar mass.  It should be noted that the assumption of these 
yields minimises the effects of the CMF, as it reduces the 
influence of very massive stars on the yields.

Models of Galactic chemical evolution in the Solar neighbourhood have 
mostly either used a Salpeter IMF with an upper mass cutoff (e.g. 
Tsujimoto et al. 1995) or a steeper function such as the Scalo 
(1986) IMF (e.g. Chiappini, Matteucci \& Gratton 
1997).  We select stars from a 
Salpeter IMF with an upper-mass cut-off of $150 M_\odot$; effective 
changes in the IMF are due solely to variations in the CMF.

\section{Results}

In Table~\ref{tab:results} we show the effect of varying the
parameters of the CMF on the number and mass of massive stars, and  on
the yields of O, Mg and Z.

As reported by Kroupa \& Weidner (2003), the slope of the high-mass
end of the IMF depends upon the form of the CMF.  The lower the mass
of a cluster, the lower the probability of forming very massive
stars.  Indeed, low-mass clusters are unable to form very high-mass
stars at all.  This can be seen in the fractions of SNII progenitors
with initial masses greater than 20, 40 and 70 $M_\odot$ ($F(>20)$,
$F(>40)$, and $F(>70)$ respectively).

\begin{figure*}
\centerline{\psfig{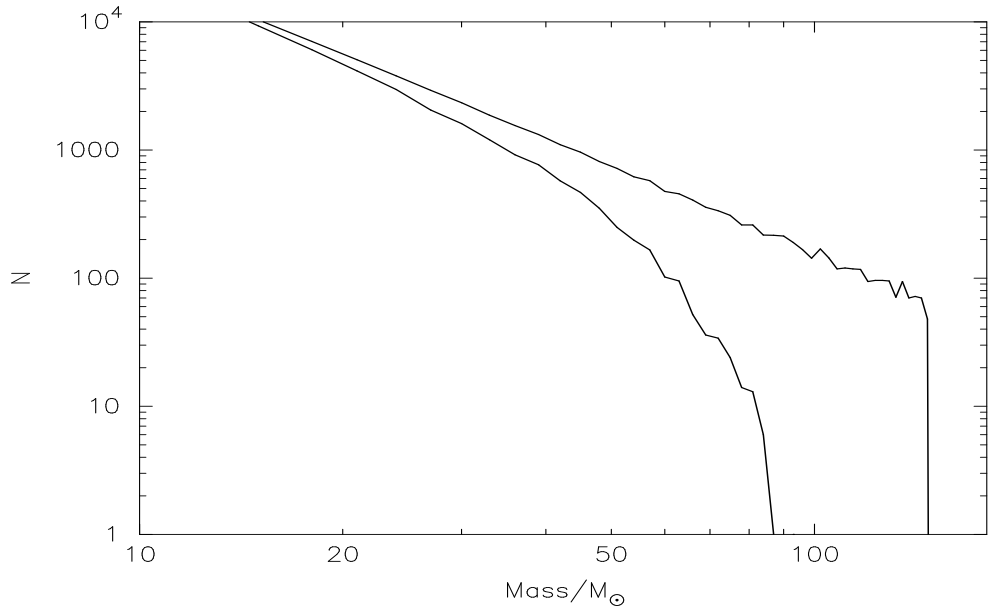}}
\caption{A comparison of the upper-end IMFs produced by CMFs with
  $(\beta,M_{\rm low},M_{\rm up})=(2,50,10^2)$ (lower curve) and
  $(2,10^3,10^6)$ (upper curve).  The $(2,10^3,10^6)$ CMF clearly
  produces far more stars at high masses.}
\label{fig:imfcomp}
\end{figure*}

In Fig.~\ref{fig:imfcomp} we compare the upper-mass IMFs of CMFs  with
($\beta$,$M_{\rm low}$,$M_{\rm up}$) = $(2,50,10^2)$ and
$(2,10^3,10^6)$.  For a CMF with parameters  $(2,50,10^2)$, which
represents a CMF that is only able to form low-mass clusters, $F(>20)
= 19\%$, $F(>40) = 3\%$ and $F(>70) = 0.1\%$.  For a maximum cluster
mass of $100 M_\odot$, it is almost impossible to form a star more
massive than a few 10s of $M_\odot$.  For a CMF with $(2,10^3,10^6)$,
all of the clusters are massive.  In this case, $F(>20) = 29\%$,
$F(>40) = 10\%$ and $F(>70) = 4\%$, which is close to what would be
expected from pure random sampling from the Kroupa (2002) IMF without the
constraint of fitting the CMF.

The mass of stars formed per SNII changes between CMFs, rising from
one per $108 M_\odot$ for $(2,50,10^2)$, to one per  $92 M_\odot$ for
$(2,10^3,10^6)$ --- a rise of $\sim 1.2$ between  the most extreme CMFs.

\begin{table*}
 \centering
 \begin{minipage}{140mm}
\caption{For a cluster mass function with slope $\beta$ and lower
and upper cut-offs  $M_{\rm low}$ and $M_{\rm up}$ respectively, we
show the mass of stars formed per SNII   $M_{\rm tot}/N_{\rm SNII}$,
the average number of SNII per cluster $N_{\rm SNII}/N_{\rm clus}$,
the fraction of SNII with masses $> 20$, 40 and $70 M_\odot$ $F(>20)$,
$F(>40)$, $F(>70)$, and the mass of oxygen,  magnesium and metals
produced per solar mass of stars formed $M_{\rm O}/M_{\rm tot}$,
$M_{\rm Mg}/M_{\rm tot}$ and $M_{\rm Z}/M_{\rm tot}$.}
\label{tab:results}
\begin{tabular} {ccccccccccc} \hline
$\beta$ & $M_{\rm low}$ & $M_{\rm up}$ & $M_{\rm tot}/N_{\rm SNII}$ &
$N_{\rm SNII}/N_{\rm clus}$ & $F(>20)$ & $F(>40)$ & $F(>70)$ & $M_{\rm
O}/M_{\rm tot}$ & $M_{\rm Mg}/M_{\rm tot}$ & $M_{\rm Z}/M_{\rm tot}$\\

 & $M_\odot$ & $M_\odot$ & $M\odot$ & cluster$^{-1}$ & \% & \%   &\%
 & $\times 10^{-3}$ & $\times 10^{-3}$ & $\times 10^{-3}$ \\ \hline

2   & 50     & $10^2$ & 108 & 0.6 & 19 & 3.1 & 0.1 & 5.4 & 0.39 & 15 \\ 
2   & 50     & $10^3$ &  97 & 1.6 & 25 & 7.0 & 1.8 & 7.2 & 0.48 & 23 \\ 
2   & 50     & $10^4$ &  95 & 2.8 & 27 & 8.6 & 2.9 & 7.7 & 0.51 & 25 \\ 
2   & 50     & $10^6$ &  93 & 5.2 & 28 & 9.4 & 3.3 & 8.1 & 0.53 & 27 \\ 
2   & $10^3$ & $10^6$ &  92 & 75  & 29 & 10  & 3.8 & 8.4 & 0.54 & 28 \\

2.5 & 50     & $10^4$ &  98 & 1.4 & 24 & 6.9 & 2.0 & 6.9 & 0.47 & 22 \\ 
1.5 & 50     & $10^4$ &  93 & 7.7 & 28 & 9.7 & 3.5 & 8.2 & 0.53 & 27 \\

\hline
\end{tabular}
\end{minipage}
\end{table*}

It is a combination of the greater number of SNII per $M_\odot$ of
stars and the change in the fractions of SNII progenitor masses that
produces a significant change in the yields between
different CMFs.  Between the $(2,50,10^2)$ and $(2,10^3,10^6)$ 
CMFs, the O, Mg and Z yields rise by a factor of 1.57, 1.40 and 
1.83 respectively.

Small differences in the yields can be seen if the slope of the CMF is
varied.  Between $(1.5,50,10^4)$ and $(2.5,50,10^4)$, the yields of O,
Mg and Z change by a factor of 1.15 - 1.23.

The total mass of stars forming was chosen to be $10^9 M_\odot$ in
order to allow a full sampling of the CMF.  If the total mass of gas
which forms stars is low, then different samplings of the CMF may have
a small effect on the yields.  For a CMF of $(2,50,10^4)$ sampled only
from $10^5 M_\odot$ of gas, the yields can vary by a factor of $\sim
1.2$.  This variation comes solely from the random sampling of the CMF
and the failure of some samples to include massive clusters and/or the
success of others at doing so.

Depending on the CMF, the number of SNII occurring per cluster changes
by a huge factor, between 0.6 for $(2,50,10^2)$, to 75 for
$(2,10^3,10^6)$.  This would be expected to play a significant role in
determining the effectiveness of the feedback of both energy and
metals into the galaxy.  When there are few SNII per cluster, it could
be expected that feedback would be relatively inefficient.  Much of
the energy from the SNII would be used destroying the natal cloud,
rather than used in spreading energy and metals far into the ISM.
When the number of SNII in a cluster is high, it may be expected that
the SNII would effectively form a superbubble, the influence of which
may be widespread.

The Solar abundance by mass of O is $\sim 6 \times 10^{-3}$, and of Mg
is $\sim 7 \times 10^{-4}$ (Lodders 2003).  The yields of oxygen are
close to Solar, while the models we have used underestimate the
yields of Mg compared to Solar.  The exact details of the yields
depend upon the models used.  Whilst it would be expected that yields
depend upon many factors such as the metallicity of the progenitor and
the details of the mass loss, we do not think that the broad
conclusions of this paper would change.  Only in the highly unlikely
case that yields scale with just the right power to offset the IMF
could the yields be totally independent of the CMF.

\section{Conclusions}

We have shown that the yields of O and Mg may vary by factors of $\sim
1.5$, and of metals by $\sim 1.8$, depending on the form of the cluster mass
function.  This is due to the inability of low-mass clusters to  form
very massive stars (cf. Kroupa \& Weidner 2003) thus changing the
fractions of contributors of different masses to enrichment.

In the Solar neighbourhood the CMF is a power-law with slope $\sim -2$
between $\sim 10^2$ and $10^4 M_\odot$ (Lada \& Lada 2003).  There is
evidence that the HII luminosity function (and so presumably the CMF)
varies with galaxy type (e.g. Kennicutt et al. 1989; Youngblood \&
Turner 1999) with early-type spirals having fewer and less bright HII
regions than late-type galaxies. This suggests a steeper CMF and/or a
lower-mass cut-off (although this conclusion is debated by Thilker et
al. 2002; their sample contains mainly late-type galaxies, however).  
The upper limit of the HII luminosity function also varies with galaxy type,
active star forming galaxies containing significantly more giant HII
regions (e.g. Alonso-Herrero et al. 2002).

It might be expected that dwarf galaxies have a CMF that has a low
upper-mass cut-off.  This is simply because dwarf galaxies have a far
smaller reservoir of gas from which to form clusters, and the most
massive clusters that they form would be expected to be fairly low.
Thus dwarf galaxies would be expected to have sub-Solar yields of
heavy elements.

The Solar Neighbourhood has a CMF that is close to ($2,50,10^4$) as shown by
Lada \& Lada (2003). This CMF produces close to the observed Solar 
abundances of O and metals, but under-produces Mg for these models.

Massive elliptical galaxies are thought to form from major merger
events (Sanders \& Mirabel 1996).  In such events, the star formation 
rate is thought to
increase dramatically and super star clusters become a dominant mode
of star formation.  Thus, a CMF with high upper and lower-mass
cut-offs may be a reasonable model for these galaxies.  In such cases
the yields could be a factor of $>1.5$ higher than in dwarf galaxies,
and $>1.1$ higher than in normal spiral galaxies due solely to
differences in the cluster mass functions.

\section*{Acknowledgments}
SPG is a UKAFF Fellow.

\label{lastpage}

\end{document}